\documentclass{article}[12pt]
\usepackage{graphicx}
\usepackage{a4}
\begin{document}
\title{A first-principles study of wurtzite-structure MnO}
\author{ P. Gopal and Nicola A. Spaldin \\ Materials Department, University of California,\\ Santa Barbara, CA 93106-5050. \\ $$\and Umesh V. Waghmare $$\\
Theoretical Sciences Unit,\\ Jawaharlal Nehru Centre for Advanced Scientific
Research,\\ Bangalore, 560 064 INDIA }

\maketitle
\begin{abstract}

We present results of a density functional theory study of MnO in the 
wurtzite structure. Our motivation is provided by recent experiments 
reporting ferromagnetism in Mn-doped wurtzite-structure ZnO. We find 
that wurtzite MnO a) is not strongly energetically disfavored compared 
with the ground state rocksalt MnO, b) shows strong magneto-structural 
coupling and c) has a piezoelectric response that is  larger than that of ZnO. 
These predictions augur well for the creation of 
ferromagnetic piezoelectric semiconductors based on Mn-doped ZnO.

\end{abstract}
\section{Introduction}
\bibliographystyle{unsrt}

The wurtzite structure adopted by ZnO does not have a center of inversion 
symmetry and therefore permits the occurrence of piezoelectricity. Indeed
previous studies have revealed that the ZnO piezoelectric coefficients are 
substantial~\cite{zno_expt}
because of incomplete cancellation between opposing ionic and electronic 
contributions \cite{ dalCorso,Hill-waghmare}. ZnO is already widely used in 
 varistors for surge protection and as filters for incoming television signals. The large piezoelectric response 
might provide improved electronic device
performance, for example, enhanced and tunable optical devices and optical wave resonators.
 
Additional interest in ZnO-based systems has recently been stimulated by 
predictions of ferromagnetism in transition metal-doped ZnO using a simple mean
field model \cite{Dietl} and also by density functional methods \cite{Sato1,
Sato2}. Numerous experimental reports of ferromagnetism \cite {ando:dms, kim,jin,
Ueda, lim} including some above room temperature \cite{nature_znmno} have
also appeared.  Although more recent experimental \cite{Risbud} and 
theoretical~\cite{Spaldin} studies call these earlier observations
into question, there is a clear need for further research in this field.

In this paper we investigate the fundamental question of what happens to the  piezoelectric response when Mn is 
substituted for Zn in the wurtzite structure. 
The answer is not obvious; some wurtzite materials, such as BeO, do not show 
a large piezoelectric response even though it is allowed by the 
non-centrosymmetricity of the crystal structure. The situation is even more 
complicated in the case of transition-metal substituents because of 
the availability 
of partially-filled $3d$ states which can become involved in chemical bonding. 
Such states have been shown to strongly suppress the occurrence of
ferroelectricity~\cite{Hill_JPC}, but their influence on piezoelectricity
has not been previously investigated. We choose 
Mn as our model transition metal dopant because the $3d$ band of the Mn$^{2+}$ 
ion is exactly half-filled, with a gap between the up-spin occupied states and 
empty down-spin states.      
Therefore we anticipate a more straightforward band 
structure than in the case of other transition metals in which one of the 
bands is usually partially filled. Also, such an arrangement is more likely
to result in an insulating state, which is a prerequisite for the occurrence
of piezoelectricity.

Our primary finding is that wurtzite MnO in fact has a larger piezoelectric 
response than ZnO by upto around  3 times. This remarkable result
suggests that, in addition to being ferromagnetic, transition metal doped 
ZnO should have a strong piezoelectric response.
Device applications that have been suggested for such magnetic piezoelectrics,
include variable transducers with magnetically-tunable piezoelectricity, and 
electric
field or stress-controlled ferromagnetic resonance devices\cite{Wood_Austin}.
Also, the ability to couple with {\it either} the magnetic {\it or}
the electric polarization will offer an extra degree of freedom in the
design of conventional sensors, actuators, transducers 
and storage devices.  The ability to transform changes in the magnetic field into electrical voltage 
and vice-versa could be  used as a magnetic field sensor for magnetic field
measurement, or  in electric current measurement. 

The remainder of this paper is organized as follows. In section II we 
summarize previous calculations on MnO in the usual rocksalt structure,
and compare our own results with those of the literature. In Sec. III 
we describe the electronic structure of MnO in the wurtzite 
structure. In Sec. IV we present the results of our studies of 
magneto-structural coupling and piezoelectricity in wurtzite MnO. 
In the last section we present our conclusions and suggestions for
future work.

\section{MnO in the Rocksalt structure} 
\label{rocksalt}

The experimental ground state crystal structure of MnO is
rocksalt~\cite{roth_mnoexpt}. In this section we present
results of our calculations for rocksalt-structure MnO in three magnetic
orderings, namely the ferromagnetic (FM) and two types of anti-ferromagnetic
(AFM) orderings. In type-I AFM ordering (AFI), the Mn ions are 
ferromagnetically aligned in the [001] planes, with adjacent planes
aligned antiferromagnetically to each other, while in type-II  
ordering (AFII) they are ferromagnetically aligned in [111] planes, again with
antiferromagnetic alignment between the planes.  
 The type-II structure
is the experimentally observed ground state. The purpose of this section 
is two-fold. First to  compare with existing data in order to verify
our methodology and pseudopotentials, and second as a point of comparison 
for our later studies on MnO in the wurtzite structure.

Our total energy calculations were carried out using the local spin density
approximation (LSDA) to density functional theory (DFT)~\cite{Kohn:dft1,Kohn:dft},
with pseudopotentials and a planewave basis as implemented in the
ABINIT~\cite{abinit} computer code. We used the optimized pseudopotentials
developed by Rappe {\it et al.},~\cite{rappe:psp} for both the Mn and
the O atoms. Our semi-core Mn pseudopotential was constructed from a $3s^2$, 
$3p^6$, $3d^5$ Mn$^{2+}$ valence configuration, with cutoff radii for the 
$s$, $p$ and $d$ states of 1.2, 1.1 and 1.4 a.u.s respectively. Optimization
of the energy cutoff was carried out using a wave vector, $q_c$, of 9 a.u.s. with
four Bessel functions. The oxygen pseudopotential used a
$2s^2$, $2p^6$ neutral valence configuration with cutoff radii of 1.2 and 1.4 a.u.
for $s$ and $p$ states respectively. Optimization of kinetic energy cutoff
for the oxygen states used $q_c$s of 7.0 and 6.5 a.u.s with four and three Bessel functions respectively. 
Good convergence was obtained with a plane-wave energy cutoff of 82 Ry
for these pseudopotentials. Rhombohedral magnetic unit cells, containing two 
formula units of MnO, were used for all three magnetic configurations. 
 Highly converged special {\bf k}-point sets consisting of 216 points in
the rhombohedral Brillouin zone were used in each case. The exchange-correlation
functional was parameterized using the Perdew-Zunger
parameterization~\cite{perdew:exc} of the Ceperley-Alder~\cite{cpley:exc1}
exchange-correlation functional.
\begin{table}
\begin{center} 
\begin{tabular}{|c|c|} 
\hline Magnetic ordering &  Energy (eV/formula unit) \\ \hline  
      AFII & 0.00 \\                   
      AF1  & 0.55 \\  
      FM    & 0.61   \\  \hline 
\end{tabular} 
\caption{Relative energies of different magnetic configurations in rocksalt-structure
MnO}
\label{MnO_energies}
\end{center}
\end{table}

First we optimized the structures for each magnetic ordering and calculated
the total energies, which are listed in Table~\ref{MnO_energies}.
As expected, we found the AFII configuration to be lowest in energy,
with the AFI 0.55 eV/formula unit higher, and the FM 0.60 eV/formula unit
above AFI. This ordering is consistent with experiment~\cite{roth_mnoexpt}, 
 first-principles Hartree Fock calculations~\cite{MnO:HF} and previous LSDA
 density functional calculations~\cite{MnO_LSDA, MnO_DFT1,MnO1_DFT}.
Our optimized rhombohedral lattice constant for AFII MnO 
was 8.16 a.u., consistent with previous LSDA calculations \cite{pask:MnO_LDA}
and showing the typical  LSDA underestimation relative to the experimental value of 8.38
a.u.~\cite{roth_mnoexpt}. 

\begin{figure}[htbp] 
\vspace{0.3in}
\begin{center}  
\includegraphics[height=4cm,width=5cm]{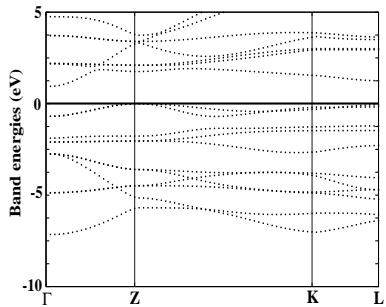} 
\end{center}
\caption{Band structure of AFII MnO plotted along the high symmetry lines of
the rhombohedral Brillouin zone.}  
\label{rocksalt_bs}
\end{figure}

In Figure \ref{rocksalt_bs}, we show our calculated band structure of AFII
MnO which is calculated at the experimental value of 8.38 a.u.  (The zero
of energy is set to coincide with the highest occupied orbital). 
The occupied {\it d } states lie at the top of the valence band, and their splitting
by the rhombohedral crystal field into two doublets and a singlet can be seen
clearly. Below and slightly separated from them are the predominantly oxygen
$p$ states which have a band width of around 4.5 eV. We will show later that the
bandwidths of the Mn and O bands are affected by both the structure and by the
magnetic ordering of the system. 
We obtain an energy  gap of 0.75 eV, in perfect agreement
with earlier LDA calculations~\cite{pask:MnO_LDA}.
Analysis of the orbital character of the LSDA band gap \cite{Filippetti_Spaldin} indicates
that it lies between predominantly up-spin $d$ states at the top of the valence band,
and down-spin $d$ states on the same Mn atom at the bottom of the conduction band. 

\section {MnO in the wurtzite structure} 

Having established that our pseudopotentials accurately reproduce previous
results for rocksalt-structure MnO, we now turn to the main part of our paper,
the study of MnO in the wurtzite structure. As mentioned in the introduction,
our motivation is provided by recent experiments in which transition metals
are doped into ZnO, where their local environment is the four-fold
coordinated structure considered here. 

\begin{figure}[htbp]
\vspace{0.2in}
\vspace{6pt}
\begin{center}
\includegraphics[height=4cm]{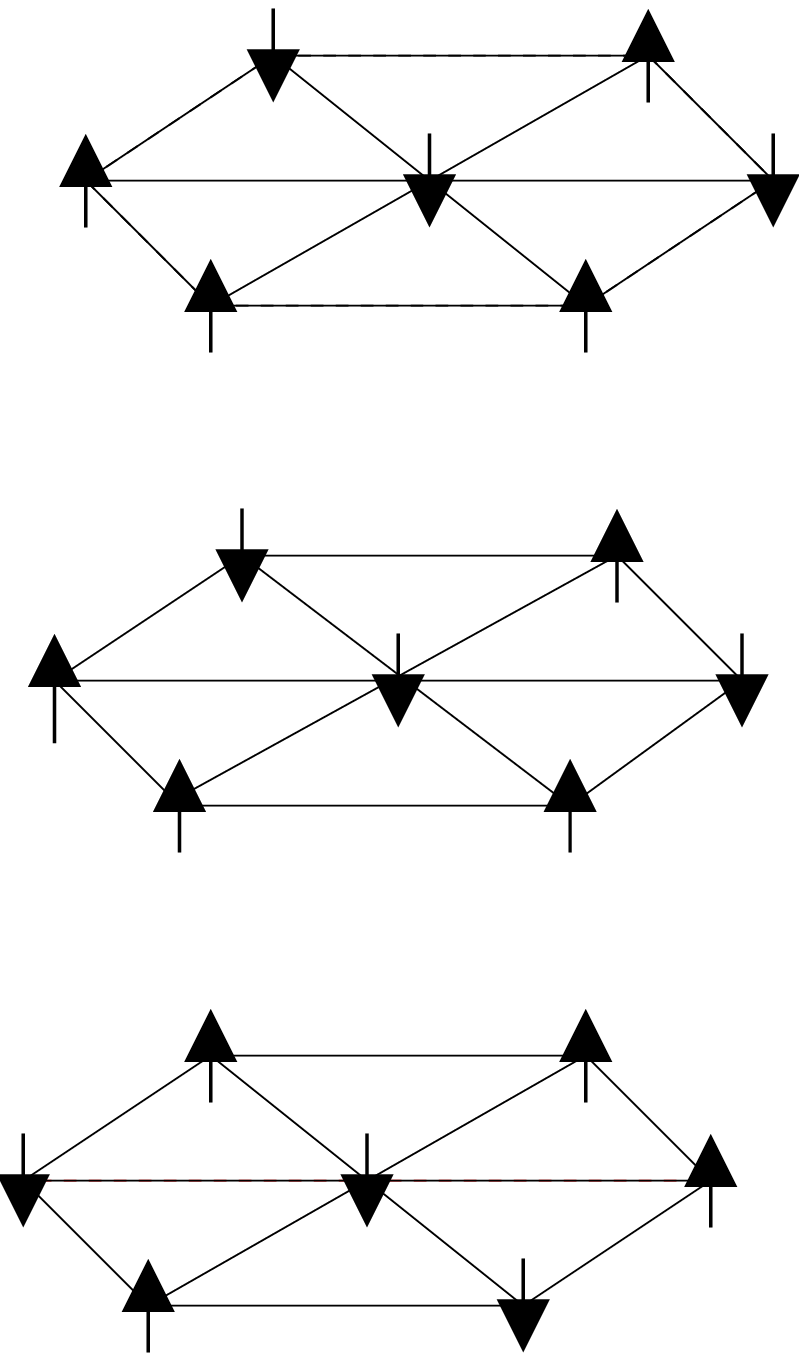} 
\hspace{3pt}
\includegraphics[height=4cm]{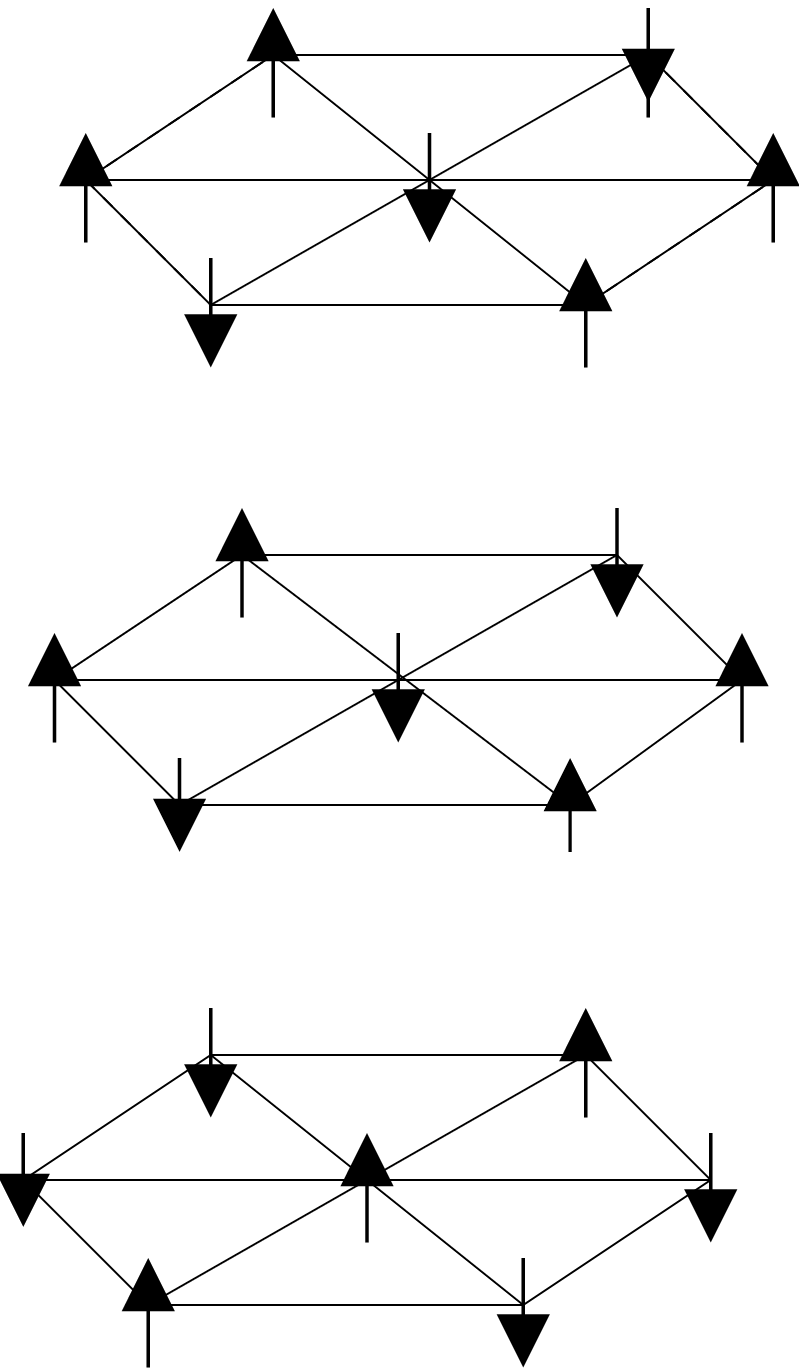}
\hspace{3pt}
\includegraphics[height=4cm]{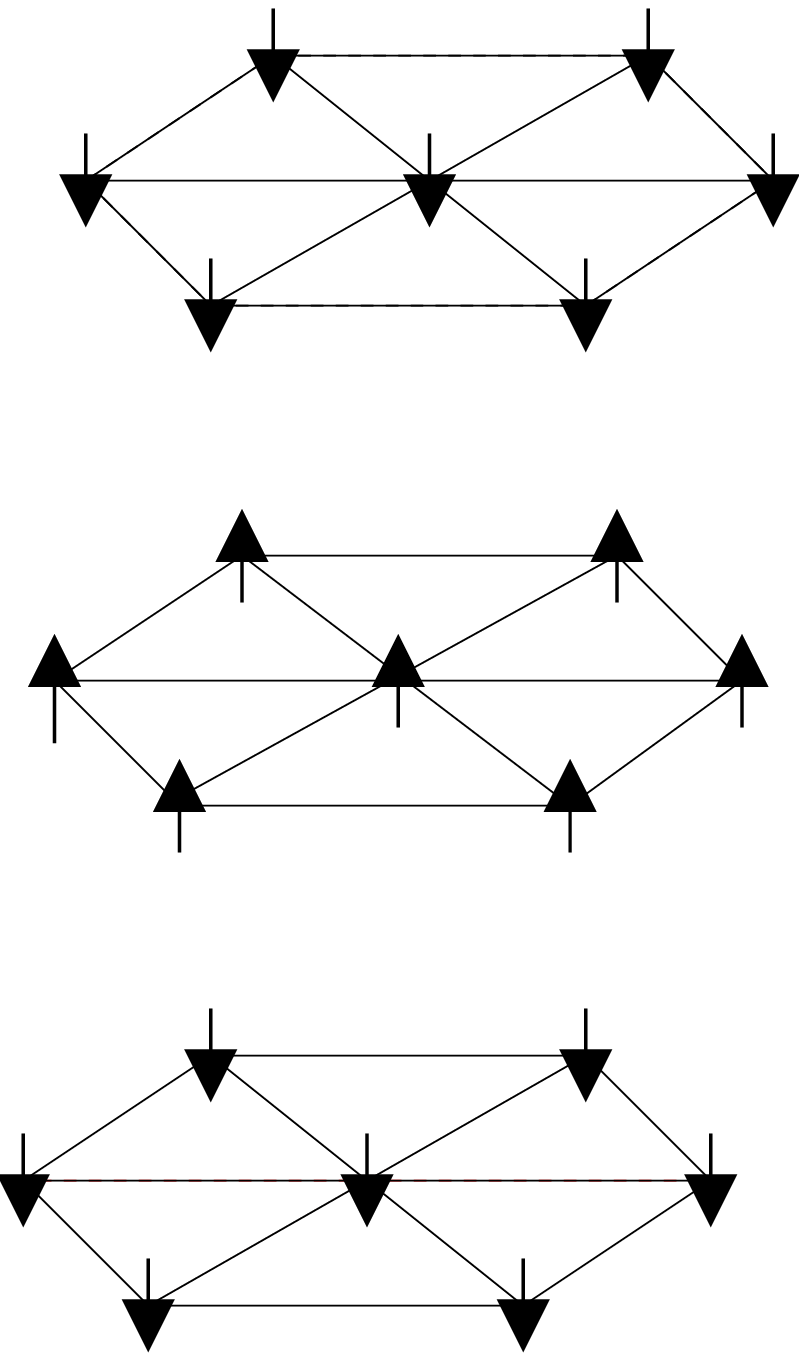}
\hspace{3pt}
\includegraphics[height=4cm]{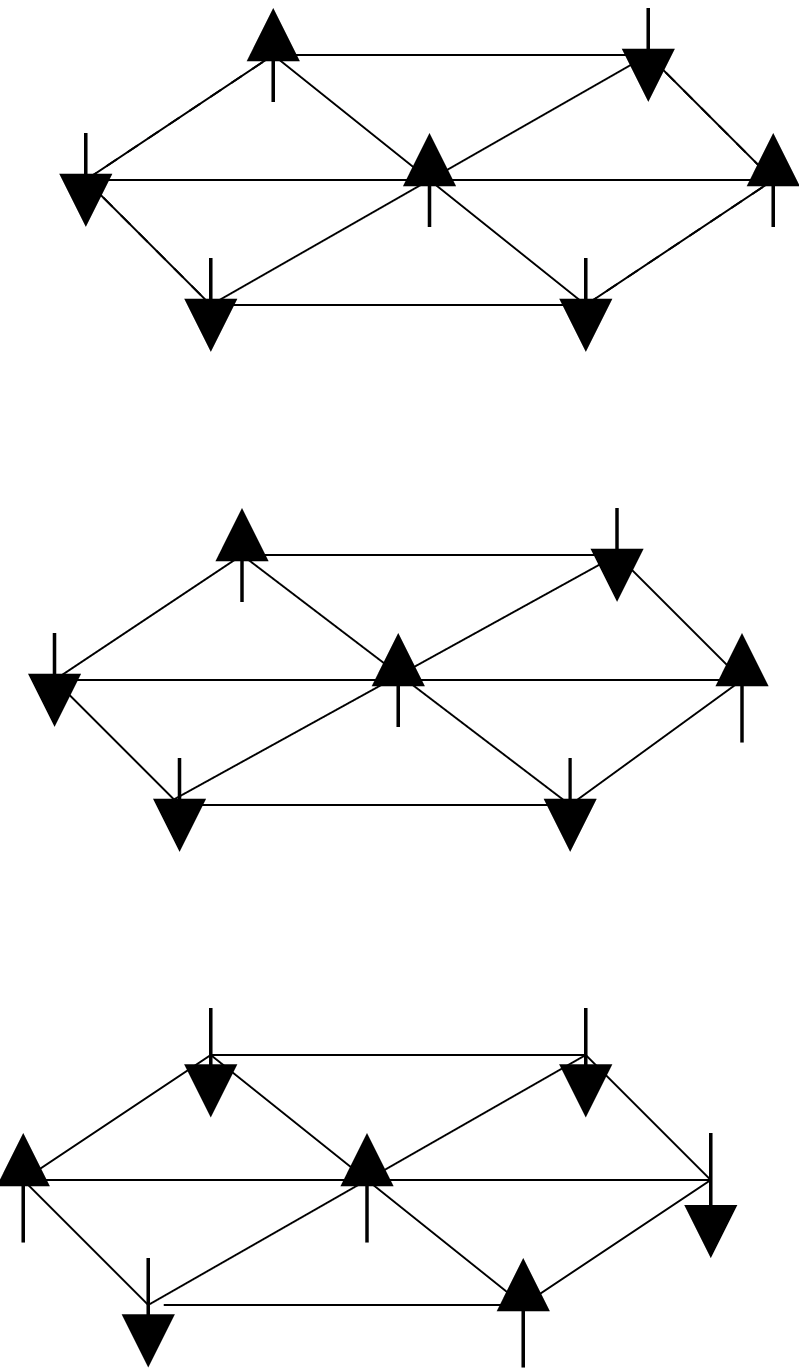}
\hspace{3pt}
\includegraphics[height=4cm]{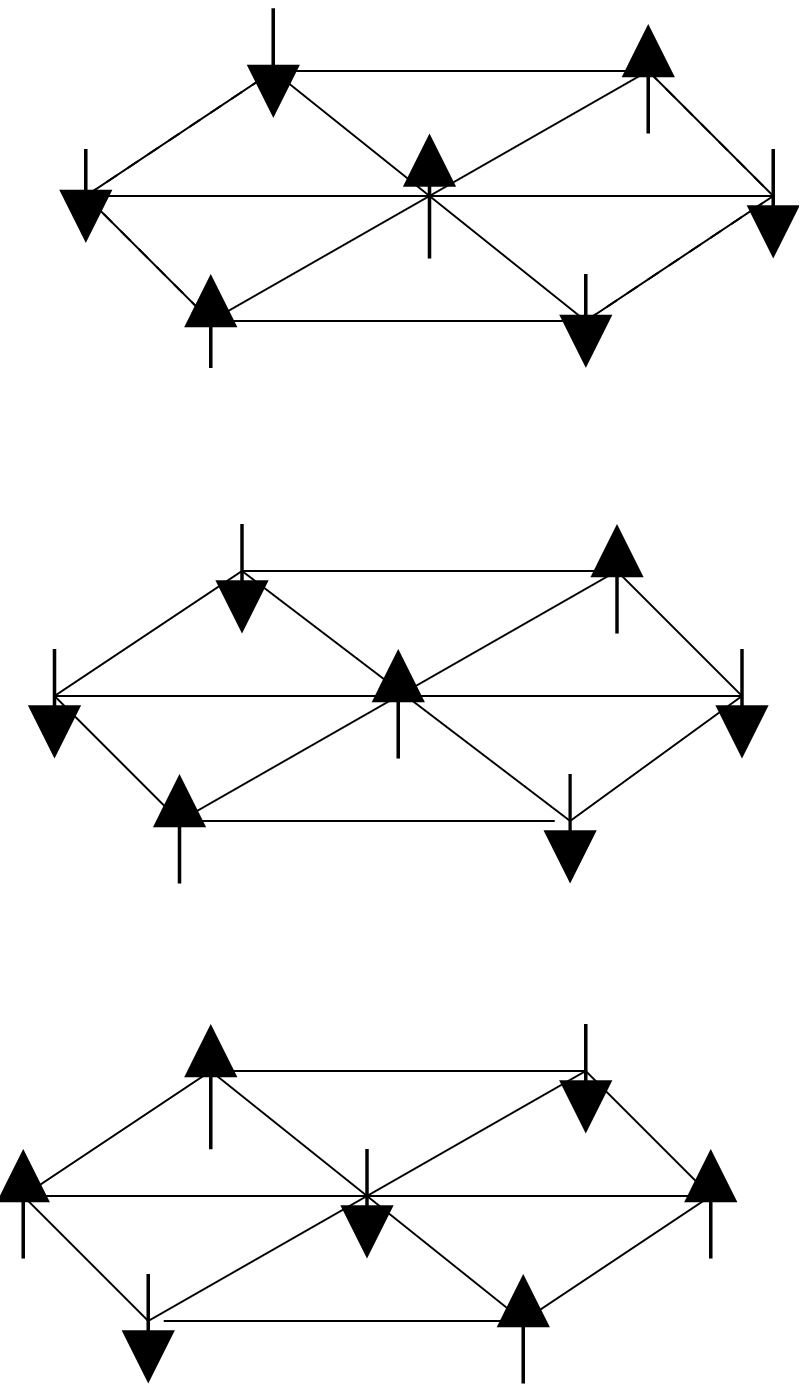}\\
\end{center}
\hspace{30pt} AF3 \hspace{50pt} AF1 \hspace{50pt} AF2b \hspace{50pt} AF4
\hspace{50pt} AF2a\\
\label{wurz_struc}

\caption{Different magnetic orderings for wurtzite structure MnO studied in this
work.}
\end{figure}

\begin{figure}[htbp] 
\vspace{0.1in}
\begin{center}  
\includegraphics[height=3.0cm]{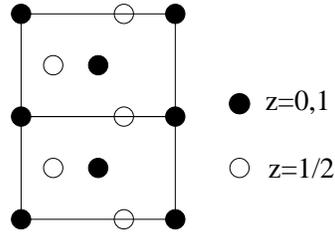} 
\end{center}
\caption{Two-dimensional projection of the unit cell of wurtzite MnO used in this work} 
\label{unit_wurz}
\end{figure}

We study the six magnetic orderings of wurtzite MnO shown in Figure~\ref{wurz_struc}.
(Note that these were previously studied for MnS 
 by Hines et al. ~\cite{hines:mn_hf}.) Identical unit cells, containing 
8 formula units, were adopted for all structures, and the pseudopotentials and cut-offs
were the same as those described in Section~\ref{rocksalt}. A $ 4 \times 4
\times 4 $ {\bf k}-point grid was used in all the calculations. A two-dimensional
projection of the unit cell, looking down the {\it c} axis is shown in Figure~\ref{unit_wurz}. The filled circles represent one set of ions in the 
the z=0 and the z=1 planes , while the empty circles show the positions of
the other ions in the  z$=\frac{1}{2}$ plane.

\section{Energetics}
 
Initially we set the lattice parameters to those of ZnO, and found that the AF3
structure was lowest in energy. However prohibitively large stresses were
obtained for all the magnetic orderings at these lattice parameters. Therefore
we chose new lattice parameters, $a$ and $c$ of 3.2847 \AA\ and 5.1789 \AA\,( around
1 \% larger than the corresponding values in ZnO) and a $u$ value of 0.3687.
These new values gave acceptably small stresses. In order to  decouple the effects of 
changes in structure from effects of changes in magnetic ordering, we then calculated
the properties of all our magnetic structures at these lattice parameters, and obtained the relative total energies 
for the wurtzite structures
 listed in Table~\ref{wurz_energies}.
 The calculated energy ordering is 

\begin{center} 
AF3 $<$ AF1 $<$ AF2b $<$ AF4 $<$ AF2a $<$ FM.
\end{center}

Note that the energies of the AF3 and AF1 phases are very similar, as
are those of the AF2b, AF4 and AF2a. We relate this below to the ferro-
and antiferromagnetic interactions between Mn nearest neighbors.
As in the rocksalt case, the ferromagnetic phase is highest in energy.
It is notable that the relative energies per formula unit of the AF3 wurtzite
structure and the AFII rocksalt structure are the same within the errors
of the calculations. This suggests that it is not unfavorable to incorporate
Mn into a tetrahedral environment. 

\begin{table}
\begin{center} 
\begin{tabular}{|c|c|} 
\hline Magnetic ordering &  Energy (eV/formula unit) \\ \hline  
AF3  &    0.000   \\  
AF1  &    0.002   \\     
AF2b &    0.098   \\ 
AF4  &    0.113   \\               
AF2a &    0.118   \\  
FM   &    0.426   \\ \hline  
\end{tabular} 
\caption{Relative energies of the different magnetic orderings in wurtzite MnO. The energies are given relative to the reference lowest energy AF3 structure.}
\label{wurz_energies}
\end{center}
\end{table}
\begin{figure}[htbp]
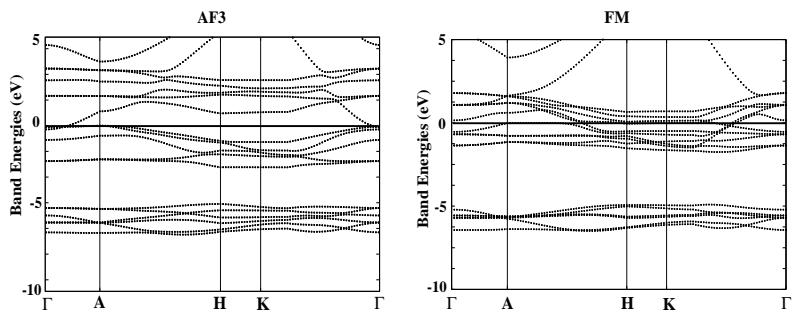
 
\vspace{0.2in}
\begin{center}  
\includegraphics[height=4cm,width=5cm]{wafm.eps} 
\hspace{5pt}
\includegraphics[height=4cm,width=5cm]{wfm.eps}
\end{center}
\caption{ Band structures of wurtzite MnO for AFM (left) and FM (right)
orderings, only the up spin states are shown in the FM case.} 
\label{wurtzite_bands}
\end{figure}

\begin{figure}[htbp] 
\begin{center}  
\includegraphics[height=3.5cm]{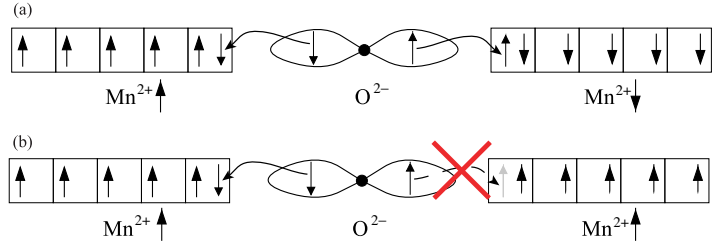} 
\end{center}
\caption{Schematic of (a) favorable Mn-O hybridization in the AFM 
configuration and (b) unfavorable hybridization in the FM configuration. In (a) the Mn ions are antiferromagnetically coupled. Therefore the down-spin
oxygen $2p$ electron undergoes covalent bond formation with an up-spin
Mn $3d$ electron (left) leaving an up-spin oxygen $2p$ electron to form
a covalent bond with a down-spin Mn $3d$ electron (right). In (b) the Mn
ions are ferromagnetically coupled. Therefore if the down-spin oxygen
$2p$ electron bonds with an up-spin Mn $3d$ electron on the left (as in
the antiferromagnetically coupled case above) the remaining up-spin
oxygen $2p$ electron is unable to bond with the {\it up-spin} Mn $3d$
electrons on the right. Therefore the amount of hybridization, and hence
the band width, is reduced in the ferromagnetic case.}
\label{cartoon}
\end{figure}

In Figure~\ref{wurtzite_bands} we show the band structures for AF3 
 and FM wurtzite structure MnO (only the up-spin states are shown in the FM
 case). Note that in both cases we obtain metallic band structures, with the FM phase having numerous
Mn $d$ derived bands crossing the Fermi energy. The AF3 phase (and the other
AFM phases, which are not shown) is close to being gapped, with only the Mn 
$4s$ band crossing the Fermi energy at the $\Gamma$ point. We point out that the
metallicity is an artifact of the underestimation of the gap by the LSDA, and in 
practice we anticipate that all of the AFM structures should be insulating. 
Again the top of the valence band is composed of predominantly Mn $3d$ states,
which this time show the $e$ - $t_2$ crystal field splitting of the tetrahedral
coordination. The predominantly oxygen $2p$ bands are again below the Mn $d$ bands and separated from them by a
gap. The bottom of the conduction band is largely composed of Mn $3d$ states,
except for the broad $4s$ band that crosses the Fermi energy. Compared with
the rocksalt band structure, the bands are narrower, as anticipated for Mn
in four-fold rather than six-fold coordination. In addition, it is notable
that the wurtzite AFM structure has broader bands than the wurtzite FM.
This reflects the increase in hybridization possible, via the usual
superexchange interaction, in the AFM bonded configuration. Mn-O hybridization
is unfavorable in the FM configuration between $d^5$ Mn$^{2+}$ ions as a 
result of Hund's rule. Cartoons of the Mn-O bonding in the AFM and FM 
cases are shown in Figure~\ref{cartoon}.
\pagebreak
\subsection{Exchange Couplings}

\begin{figure}[htbp]
\vspace{0.01in}
\vspace{2pt}
\begin{center}
\includegraphics[height=4cm]{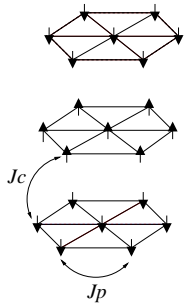} 
\caption{Schematic of the exchange parameters $J_p$ and $J_c$.}
\label{Jfig}
\end{center}
\end{figure}

To understand the relative energies of the different spin orderings, we 
next analyze the energy differences between the configurations using a  
two-parameter Heisenberg model; 
\begin{equation}
 H = E_0 + \sum J_{ij} S_i. S_j
\end{equation} 
where $S_i.S_j$ =+1 if the spins $i$ and $j$ are parallel,
and  -1 if they are antiparallel. We include two different 
values for $J$; an in-plane Mn-Mn interaction,
$J_p$, and an out-of-plane (along $c$) interaction, $J_c$, as shown
in Figure~\ref{Jfig}. Note that, for the ideal wurtzite structure, the
in-plane and out-of-plane Mn-Mn distances are identical, and we would
expect to obtain $J_p = J_c$ in the absence of any kind of spin-orbit coupling. However, we constrain the parameter u to be 0.368 which is smaller than the ideal value of 0.375 and hence we find a percentage difference in $J_p$ and $J_c$ of the order of the deviation from the ideal value.  

\begin{center}
\begin{table}
\begin{center}
\begin{tabular}{|c|c|c|c|c|c|c|c|}
\hline
& & AF3&AF1&AF2b&FM&AF4&AF2a\\
\hline
in-plane & $\uparrow \downarrow $& 4 & 4& 0&0 &4&4\\
           & $\uparrow \uparrow$ & 2 & 2&6&6&2&2\\
\hline
out-of-plane& $\uparrow \downarrow$ & 4 &4&6&0&2 & 2 \\
            & $\uparrow \uparrow $ & 2 &2 & 0&6 & 4& 4\\
\hline
\end{tabular}
\caption{Number of in-plane and out-of-plane FM and AFM interactions
for each magnetic configuration.}
\label{JTable}
\end{center}   
\end{table}
\end{center}  
Table~\ref{JTable}  gives the number of parallel and antiparallel in-plane 
and out-of-plane nearest neighbors for a Mn ion in each magnetic structure.
Note that the AF3 and AF1 structures have identical types of nearest-neighbor
(nn) Mn-Mn interactions (4 in-plane AFM and 4 out-of-plane AFM), which explains their similar energies. Similarly, AF4 and AF2a have identical numbers of nn
interactions to each other, but have fewer AFM interactions than AF3 and AF1 (4 in-plane AFM and 2 out-of-plane AFM). AF2b,
which is close in energy to AF4 and AF2a has the same {\it total} number
of nn AFM interactions (six), but they all occur out-of-plane. Finally the
FM phase has no AFM interactions, leading to its correspondingly higher
energy.  
By fitting the energies of the six configurations onto the Heisenberg
model, we obtained the parameters $J_p = -0.023$ eV and $J_c = -0.028$ eV, 
with a root mean square deviation of 0.003. The small deviation indicates 
that a nearest-neighbor Heisenberg model describes the energetics of the
system well, and suggests that more distant exchange interactions are weak
compared to the nearest-neighbor interactions. Since the Mn-Mn in-plane bond
length is almost equal to the out-of-plane Mn-Mn bond length, we obtain
$J_p \approx J_c$ as expected.

To conclude this section, we note that AFM coupling is more favourable than FM
for both in-plane and out-of-plane Mn-O-Mn bonds and. This suggests that, if
doping of Mn ions into ZnO results in clustering of the magnetic ions, we will
obtain strong antiferromagnetism. 

\subsection{Forces and stresses}

In many magnetic compounds, a change of magnetic ordering causes a stress
which induces a structural distortion \cite{alessio:crn}. For example, in 
rocksalt MnO, AFM spin ordering causes a reduction in magnetic symmetry from 
cubic to rhombohedral, which is accompanied by a structural 
transition~\cite{massida:mn}. In some materials, in which there is 
strong magneto-structural coupling, the distortion
can be quite large. For example in CrN the magnetic ordering induces 
a transition from cubic to orthorhombic
symmetry~\cite{alessio:crn,CrN_expt1,CrN_expt2}.
And in LaMnO$_3$ the observed A-type antiferromagnetic ordering is unstable in the presence of distortions~\cite{Pickett_singh}.
The concept of {\it magnetic stress} has been introduced to describe 
structural phase transitions that are induced by magnetic
ordering~\cite{alessio:crn}. In this section we investigate the magneto-structural coupling in 
wurtzite MnO, as a precursor to studying the piezoelectric properties
in the next Section.

Keeping the lattice parameters of all the structures fixed at the AF3 values, the spin ordering was changed to each of the other five configurations in turn, and the electronic structure was obtained and the stresses and the forces for each magnetic phase were calculated. In the table below, the  relative stresses 
on the unit cell along the $z$ direction and largest  forces on the Mn 
atoms are tabulated for each magnetic ordering. 
\vspace{10pt}
\begin{center}
\begin{tabular}{|c|c|c|c|c|c|}
\hline
 & AF3&AF1&AF2b&AF4&AF2a\\
\hline
Stress $\sigma_{33}$ &-7.35 $\times 10^{-5}$ & -7.067 $\times 10^{-5}$ 
   & 8.35 $\times 10^{-5}$& -2.39 $\times 10^{-4}$ & -2.46 $\times 10^{-4}$\\
 (Ha/bohr$^3$) & & & & & \\
\hline
Force(max) & $5.87 \times 10^{-1} $  &5.76 $\times 10^{-1}$ & 4.42 $\times 10^{-3}$& 8.43 $\times 10^{-1}$&8.24 $\times 10^{-1}$  \\
 (Ha/bohr) & & & & & \\
\hline
\end{tabular}
\end{center}
\vspace{0.25in}
We observe similar groupings as those obtained for the relative energies. 
The AF1 and AF3 phases are similar in the spin-environment and therefore have similar stresses and forces for the same structure. Likewise, the 
 AF2a and AF4 spin orderings show similar stresses and forces to each other, but both are considerably larger than those in the AF3 and AF1 structures. The  significant change in the 
stress when the magnetic ordering is changed indicates a strong 
coupling between the magnetic moments of the atoms and the crystal structure. 

\section{Piezoelectricity}

The piezoelectric constant relates the stress response of a material to a
uniform electric field or conversely  the macroscopic polarization induced by a 
a macroscopic strain. The latter definition is used here to calculate the
electro-mechanical response from first principles. We calculated the changes in 
macroscopic polarization  using the Berry phase approach~\cite{vander:pol}
within the ABINIT density functional program~\cite{abinit}. First, the
equilibrium structure parameters ($a_0$, $c_0$ and $u_0$) were obtained for each
magnetic configuration, in turn and the spontaneous polarizations, $P_0$, (along
the z-axis) were calculated. 
Then the 
 reference structure was strained along the z axis by $1 $ \%  
($e_{zz}=0.01$, $e_{zz}$ is the percentage change in strain along the z-axis ) and its structure relaxed with respect to the internal strain (structural parameter $u$). The polarization $P_1$ for the relaxed structure was then calculated, and the  
piezoelectric co-efficient $e_{33}$ was obtained using its
definition:
\begin {equation} 
e_{33} = \frac {(P_1 -P_0)}{e_{zz}}
\end{equation} 
Subsequent application of a -1\% strain along the z-axis showed that the
polarization deviated slightly from linearity in MnO (in contrast to ZnO, where
it is strictly linear over that range with the same set of convergence
parameters.)
All the Berry phase polarization calculations used a 2$\times$2$\times$8
\textbf{k}-point mesh and a wavefunction tolerance set to 10$^{-15}$. Calculations for one of the configurations were repeated
with a 3$\times$3$\times$8 mesh to test convergence. A 2$\times$2 mesh 
in the ab-plane is found to be sufficient as it corresponds to a
4$\times$4 mesh for the primitive unit cell of the wurtzite structure. In order
to avoid the $\Gamma$ point, where the gap vanishes indicating metallicity, the
\textbf{k}-point grid was shifted by $\frac{1}{2}$ along the three co-ordinate
axes. 

 Our calculated values of $e_{33}$ for all the five magnetic orderings are listed in
Table 3. While we find groupings of configurations (AF3 and AF1), (AF4 and AF2a) such that the energies and stresses
are similar for members of a given group, their piezoelectric constants are found to
be somewhat dissimilar, as they are second order derivatives of energy.

The piezoelectric constant of wurtzite MnO in the lowest energy 
AF3 configuration is about 3 times larger than that of ZnO
$(1.0-1.2 C/m^2$\cite{Hill-waghmare}). In ZnO, the phonon (internal strain)
contribution to the piezoelectric constant was found to  dominate and this 
was attributed partly to hybridization between the filled 3d band of Zn and the
2p band of oxygen.
 In AF3 MnO, we find that the electronic contribution
to $e_{33}$ is also significant and in fact has the same sign as that of
the phonon contribution (in contrast with ZnO, where the two contributions partially compensate). The large electronic contribution arises because the $d$ band of Mn is only partially occupied and  can therefore contribute directly to the piezoelectric response through change in its electron content. 

Note that the enhanced piezoelectric response which we find in wurtzite MnO is
consistent with earlier observations of large dielectric and piezoelectric
constants in systems that are close to electronic and structural instabilities
\cite{Waghmare-Hill-Seshadripaper}. Wurtzite structure MnO is both unstable structurally (with the rocksalt structure being the grounds state ) and has other magnetic configurations close in energy. 
Finally we point out that Mn doped ZnO should have a strong piezoelectric response provided that it remains insulating.
\vspace{20pt}
\begin{table}
\begin{center}
\begin{tabular}{|c|c|c|c|c|c|}
\hline
Ordering & AF3 &AF1 &AF2b &AF4 & AF2a  \\
\hline
$e_{33}$(C/m$^2$) & 4.81 & 11.24 & 0.911 & 2.05 &1.30 \\
\hline
$P_{0}$(C/m$^2$) & 0.045&0.053&0.017& 0.014&-0.072\\
\hline
\end{tabular}
\label{polartable}
\caption{Piezoelectric constant, $e_{33}$ and the spontaneous polarization,$P_0$, $(C/m^2)$ for the different magnetic orderings of wurtzite MnO.}
\end{center}
\end{table}

\section{Summary and Conclusions}

In this research we investigated the electronic properties of wurtzite MnO 
using  density functional calculations. Using Berry phase polarization 
calculations we found that magnetic MnO is a stronger piezoelectric than non-magnetic
 ZnO. This to our knowledge is the first calculation of piezoelectric response in
a magnetic material and shows clearly that presence of magnetism does not
preclude a piezoelectric response and can even enhance it. 
The reason for the enhancement is that, in addition to the phonon contribution found in ZnO, the electrons
 in the partially filled $d$ band of Mn also contribute to the piezoelectricity. Our results suggest that 
 piezoelectricity should persist in Mn doped ZnO provided that it remains insulating.
   On analyzing the energetics of the different
magnetic orderings for wurtzite MnO, however we find that anti-ferromagnetic
ordering is more stable than  ferromagnetic ordering with the most stable
structure having the largest number of nearest neighbor antiferromagnetic
interactions. From this we can
conclude that if doping of Mn in ZnO results in clustering, ferromagnetism
is unlikely. 

\section{Acknowledgements}
 This work was supported by the Department of Energy, grant number
DE-FG03-02ER45986, and made use of MRL central facilities, supported
by the the National Science Foundation under the
Award No. DMR00-80034. UVW thanks the Materials Department at the University of California, Santa Barbara for their hospitality.

\bibliography{paper1} 

\begin{thebibliography}{10}

\bibitem{zno_expt}
I.B. Kobiako.
\newblock {\em Solid. State. Commun.}, 35:305, 1979.

\bibitem{dalCorso}
A.~Dal Corso, M.~Posternak, R.~Resta, and A.~Baldereschi.
\newblock {\em Phys. Rev. B.}, 50:10715, 1994.

\bibitem{Hill-waghmare}
N.A. Hill and U.V. Waghmare.
\newblock {\em Phys. Rev. B.}, 62:8802, 2002.

\bibitem{Dietl}
T.~Dietl, H.~Ohno, F.~Matsukura, J.~Cib\`ert, and D.~Ferrand.
\newblock {\em Science}, 287:1019, 2001.

\bibitem{Sato1}
K.~Sato and H.~Katayama-Yoshida.
\newblock {\em Jpn. J. Appl. Phys.}, 39, 2000.

\bibitem{Sato2}
K.~Sato and H.~Katayama-Yoshida.
\newblock {\em Phys. Stat. Sol B}, 229:673, 2002.

\bibitem{ando:dms}
K.~Ando, H.~Saito, Z.~Jin, T.~Fukumura, Y.~Matsumoto, and A.~Ohtomo.
\newblock {\em Appl. Phys. Lett.}, 78:2700, 2001.

\bibitem{kim}
J.H. Kim, H.~Kim, D.~Kim, Y.E. Ihm, and W.K. Choo.
\newblock {\em Physica B}, 327:304, 2003.

\bibitem{jin}
Z.W. Jin, T.~Fukumura, K.Hasegawa, Y.Z. Yoo, K.Ando, T.~Sekiguchi, P.Ahmet,
  T.Chikyow, T.~Hasegawa, H.~Koinuma, and M.~Kawasaki.
\newblock {\em J. Crystal Growth}, 237:548, 2002.

\bibitem{Ueda}
K.~Ueda, H.~Tabata, and T.~Kawai.
\newblock {\em Appl. Phys. Lett.}, 79:988, 2001.

\bibitem{lim}
S.W. Lim, D.K. Hwang, and J.M. Myoung.
\newblock {\em Solid. State. Commun.}, 125:231, 2003.

\bibitem{nature_znmno}
Parmanand Sharma, Amita Gupta, Frank J.~Owens K.V.~Rao, Renu Sharma, Rajeev
  Ahuja, J.M.~Osorio Guillen, {B\" orje} Johansson, and G.A. Gehring.
\newblock {\em Nature Materials}, 2:673, 2003.

\bibitem{Risbud}
A.S. Risbud, N.A. Spaldin, Z.Q. Chen, S.~Stemmer, and R.~Seshadri.
\newblock {\em Phys. Rev. B.}, 68:205202, 2003.

\bibitem{Spaldin}
Nicola Spaldin.
\newblock {\em cond-mat/0306477}.

\bibitem{Hill_JPC}
N.A. Hill.
\newblock {\em J. Phys. Chem. B}, 104:6694.

\bibitem{Wood_Austin}
V.~E. Wood and A.~E. Austin.
\newblock {\em in Magnetoelectric interaction phenomena in crystals}.
\newblock Gordon and Breach, 1975.

\bibitem{roth_mnoexpt}
W.L. Roth.
\newblock {\em Phys. Rev. B.}, 110:1333, 1958.

\bibitem{Kohn:dft1}
P.~Hohenberg and W.~Kohn.
\newblock {\em Phys. Rev.}, 136:864, 1964.

\bibitem{Kohn:dft}
W.~Kohn and L.J. Sham.
\newblock {\em Phys. Rev.}, 140:1133, 1965.

\bibitem{abinit}
X.~Gonze \textit{et al}.
\newblock {\em ABINIT is a plane wave pseudopotential, a common project of the
  Universite Catholique de Louvaiun, Corning Incorporated and other
  contributors (URL http://www.abinit.org).}

\bibitem{rappe:psp}
A.M. Rappe, K.M. Rabe, E.~Kaxiras, and J.D. Joannopolous.
\newblock {\em Phys. Rev. B.}, 41:1227, 1990.

\bibitem{perdew:exc}
J.P. Perdew and A.~Zunger.
\newblock {\em Phys. Rev. B.}, 23:5048, 1981.

\bibitem{cpley:exc1}
D.M. Ceperley and B.J. Alder.
\newblock {\em Phys. Rev. Lett.}, 45:566, 1980.

\bibitem{MnO:HF}
W.C. MacKrodt and E.A. Williamson.
\newblock {\em J.Phys. Condens. Matter.}, 9:6591, 1997.

\bibitem{MnO_LSDA}
K.~Terakura, A.R.~Williams T.~Oguchi, and J.~Kubler.
\newblock {\em Phys. Rev. B.}, 30:4734, 1984.

\bibitem{MnO_DFT1}
L.F. Mattheiss.
\newblock {\em Phys. Rev. B.}, 5:290, 1972.

\bibitem{MnO1_DFT}
S.~Massida, M.~Posternak, A.~Baldereschi, and R.~Resta.
\newblock {\em Phys. Rev. Lett.}, 82:430, 1999.

\bibitem{pask:MnO_LDA}
J.E. Pask, D.J. Singh, and I.I. Mazin.
\newblock {\em Phys. Rev. B.}, 64:024403, 2001.

\bibitem{Filippetti_Spaldin}
A.~Filippetti and N.~A. Spaldin.
\newblock {\em Phys. Rev. B.}, 67:125109, 2003.

\bibitem{hines:mn_hf}
R.I. Hines, N.L. Allan, G.S. Bell, and W.C. Mackrodt.
\newblock {\em J.Phys. Condens. Matter.}, 9:7105, 1997.

\bibitem{alessio:crn}
A.~Filippetti and N.A. Hill.
\newblock {\em Phys. Rev. Lett.}, 85:5166, 2000.

\bibitem{massida:mn}
S.~Massidda {\it et al}.
\newblock {\em Phys. Rev. Lett.}, 82:430, 1999.

\bibitem{CrN_expt1}
L.M. Corliss, N.~Elliott, and J.M. Hastings.
\newblock {\em Phys. Rev.}, 117:929, 1960.

\bibitem{CrN_expt2}
P.~Subramanya~Herle {\it et al,}.
\newblock {\em J. Solid. State Chem.}, 134:120, 1997.

\bibitem{Pickett_singh}
D.~J. Singh and W.~E. Pickett.
\newblock {\em Phys. Rev. B.}, 57:88, 1998.

\bibitem{vander:pol}
R.D. King-Smith and D.~Vanderbilt.
\newblock {\em Phys. Rev. B.}, 47:1651, 1993.

\bibitem{Waghmare-Hill-Seshadripaper}
U.~V. Waghmare, N.~A. Hill, H.~Kandpal, and R.~Seshadri.
\newblock {\em Phys. Rev. B.}, 67:125111, 2003.

\end{thebibliography}

\end{document}